\documentstyle[preprint,aps]{revtex}
\begin{document}
\preprint{}
\draft

\title{
Stopping and Radial Flow in Central $^{58}$Ni + $^{58}$Ni Collisions\\
between 1 and 2 AGeV
}

\author{
B.~Hong$^{4}$\footnote{Present address: Korea University, Seoul, South Korea}, 
N.~Herrmann$^{4,6}$, J. L.~Ritman$^{4}$, D.~Best$^{4}$, 
A.~Gobbi$^{4}$, K.~D.~Hildenbrand$^{4}$, M.~Kirejczyk$^{4,10}$, 
Y.~Leifels$^{4}$, C.~Pinkenburg$^{4}$, W.~Reisdorf$^{4}$, D.~Sch\"ull$^{4}$, 
U.~Sodan$^{4}$, G.~S.~Wang$^{4}$, T.~Wienold$^{4}$, J.~P.~Alard$^{3}$, 
V.~Amouroux$^{3}$, 
N.~Bastid$^{3}$, I.~Belyaev$^{7}$, G.~Berek$^{2}$, J.~Biegansky$^{5}$, 
A.~Buta$^{1}$, 
J.~P.~Coffin$^{9}$, P.~Crochet$^{9}$, R.~Dona$^{9}$, P.~Dupieux$^{3}$, 
M.~Eskef$^{6}$, P.~Fintz$^{9}$, Z.~Fodor$^{2}$, L. Fraysse$^{3}$,
A.~Genoux-Lubain$^{3}$, G.~Goebels$^{6}$, G.~Guillaume$^{9}$, 
E.~H\"afele$^{6}$, 
F.~Jundt$^{9}$, J.~Kecskemeti$^{2}$, M.~Korolija$^{6}$, R.~Kotte$^{5}$, 
C.~Kuhn$^{9}$, A.~Lebedev$^{7}$, I.~Legrand$^{1}$, C.~Maazouzi$^{9}$, 
V.~Manko$^{8}$, J.~M\"osner$^{5}$, S.~Mohren$^{6}$, W.~Neubert$^{5}$, 
D.~Pelte$^{6}$, M.~Petrovici$^{1}$, P. Pras$^{3}$, F.~Rami$^{9}$, 
C.~Roy$^{9}$, Z.~Seres$^{2}$, B.~Sikora$^{10}$, V.~Simion$^{1}$, 
K.~Siwek-Wilczy\'{n}ska$^{10}$, A.~Somov$^{7}$, L.~Tizniti$^{9}$, 
M.~Trzaska$^{6}$, M.~A.~Vasiliev$^{8}$, P.~Wagner$^{9}$, 
D.~Wohlfarth$^{5}$, A.~Zhilin$^{7}$\\
(FOPI Collaboration)
}

\address{
$^1$ Institute for Nuclear Physics and Engineering, Bucharest, Romania\\
$^2$ Central Research Institute for Physics, Budapest, Hungary\\
$^3$ Laboratoire de Physique Corpusculaire, IN2P3/CNRS, and Universit\'{e}
Blaise Pascal, Clermont-Ferrand, France\\
$^4$ Gesellschaft f\"ur Schwerionenforschung, Darmstadt, Germany\\
$^5$ Forschungszentrum Rossendorf, Dresden, Germany\\
$^6$ Physikalisches Institut der Universit\"at Heidelberg, Heidelberg, Germany\\
$^7$ Institute for Theoretical and Experimental Physics, Moscow, Russia\\
$^8$ Kurchatov Institute, Moscow, Russia\\
$^9$ Centre de Recherches Nucl\'{e}aires and Universit\'{e} Louis Pasteur,
Strasbourg, France\\
$^{10}$ Institute of Experimental Physics, Warsaw University, Poland\\
}

\maketitle

\begin{abstract}
The production of charged pions, protons and deuterons has been studied 
in central collisions of $^{58}$Ni on $^{58}$Ni at incident beam energies 
of 1.06, 1.45 and 1.93 AGeV. The dependence of transverse-momentum and
rapidity spectra on the beam energy and on the centrality of the collison
is presented.
It is shown that the scaling of the mean
rapidity shift of protons established for AGS and SPS energies is valid
down to 1 AGeV.
The degree of 
nuclear stopping is discussed; the IQMD transport model
reproduces the measured proton rapidity spectra for the most central events
reasonably well, but does not show any sensitivity between the soft and the 
hard equation of state (EoS).
A radial flow analysis, using the midrapidity 
transverse-momentum spectra, delivers freeze-out temperatures 
$T$ and radial flow velocities $\beta_{r}$ which increase with 
beam energy up to 2 AGeV; in comparison to existing data
of Au on Au over a large range of energies only $\beta_{r}$ shows a 
system size dependence.
\end{abstract}
\vspace{1.0cm}


\section{introduction}

One of the main topics of the current relativistic heavy ion 
experiments is the determination of the properties of nuclear
matter at high densities and temperatures~\cite{rstock1,hsto1,gb1}. 
Many interesting effects can happen 
already at densities just twice normal nuclear matter density
that can be reached in collisions at incident energies 
around 1~AGeV:
The masses of the constituents can be affected by the surrounding medium
due to chiral symmetry restoration~\cite{gb1,lutz1,gqli1,aich1}, 
a sizable fraction of the constituents can be excited 
into hadronic resonances whose 
lifetime and mutual interactions might be different 
in comparison to the properties of the free particles,
and for a short time the system might form an equilibrated state that 
could give access to the bulk properties of nuclear matter, i.e. to
the fundamental nuclear matter equation of state 
(EoS)~\cite{gqli1,skapu1,rstock2,hari1}.

Despite many efforts the various effects could not
be disentangled so far, although some fascinating observations 
have been made recently that await a consistent and coherent explanation,
e.g. large collective flow~\cite{wr1,mal1}, 
low entropy production~\cite{ck1}, reduced pion multiplicities in heavy systems
with respect to small systems~\cite{pelte1}, 
and enhanced subthreshold kaon production in central collisions\cite{kaos1}.
In order to achieve an understanding of the underlying physics those 
observations that were 
obtained at various beam energies and with different 
systems need to be correlated with each other.
This can be done a) by studying 
different observables in one system under the same conditions,
b) by comparing the same observable for a variety of incident energies
and system sizes. 
The latter point is particularly interesting since 
some observables became available from experiments at the much higher
beam energies of the BNL AGS and the CERN SPS, where changes 
in the development of certain observables could be caused by a
possible phase transition to the QCD deconfinement state~\cite{qm96}.
Hence a consistent comparison over more than two orders of 
magnitude in beam energy becomes possible.

A prerequisite for the understanding of the 
high density and temperature phase of nuclear matter
that might have prevailed in the initial stage of the reaction,
is the knowledge and the description of the final state 
that reflects the properties when the constituents cease to 
interact (freeze-out). The freeze-out conditions visible in the distributions 
of hadrons are very important since they allow to
test concepts like equilibrium and stopping, and therefore 
are very useful as a constraint for all more elaborated theories.

This paper presents a rather complete set of data for the phase space 
distribution (transverse momentum $p_{t}$ and rapidity $y$) 
of protons and deuterons as well as pions for the central collisions 
of $^{58}$Ni on $^{58}$Ni at incident beam energies between 1 and 2 AGeV, 
a system for which 
pion~\cite{dp1}, kaon~\cite{jrit1} and $\Delta$ resonance~\cite{bhong1} 
production as well as proton-proton correlations ~\cite{rk1} 
have already been studied.
In the following
the centrality and beam energy dependence of the momentum space 
distributions of the most abundant hadrons are discussed in detail. 
They provide the basic requirements which models need to fulfill, 
before the discussion of the initial temperature or baryon density can 
start. IQMD transport model~\cite{bass1} calculations are compared to the 
experimental proton and deuteron rapidity distributions. 
This allows to address 
the question of nuclear stopping power, although in a model\,-\,dependent way. 
We present the mean rapidity shift of protons 
and study whether the known scaling behavior established 
above 10 AGeV~\cite{fv1} is satisfied at lower beam energies. 
Finally, we interpret the midrapidity 
data in terms of a thermal model including collective radial flow.
In order to facilitate comparison with other data
we use the simple assumption proposed by Siemens and Rasmussen~\cite{ps1}, 
although a lot of effort is devoted to the development and application of more 
realistic expansion scenarios~\cite{wr1,pd1,mpt1}. 

\section{experiment}

The experiment was performed at the heavy ion synchrotron SIS at GSI
by bombarding $^{58}$Ni beams of 1.06, 1.45 and 1.93 AGeV on a fixed $^{58}$Ni
target of 225 mg/cm$^{2}$ (about 1$\%$ interaction length), placed
in the target position of the FOPI detector, which is described
in detail elsewhere~\cite{ag1,jr1}. 
For the analysis presented in this 
paper we used the central drift chamber (CDC) of FOPI for particle 
identification, and its forward plastic wall for the centrality determination. 
This azimuthally symmetric forward wall covers the polar  
angles $\theta_{L}$ from 7$^{o}$ to 30$^{o}$, measuring the deposited 
energy and the time of flight and hence the charge of the fragments.
The multiplicity of these fragments, PMUL, was used for the selection 
of the event centrality. The CDC covers the $\theta_{L}$ angles
from 30$^{o}$ to 150$^{o}$. Pions, protons and deuterons were 
identified in the chamber by means of their 
mean energy loss $<dE/dx>$ and their laboratory momentum $p_{L}$,
obtained from the curvature of the particle tracks
in the field of a 0.6~T magnet.
The accuracy of the position measurements of the tracks 
in radial and azimuthal direction via drift time was
$\sigma_{r \phi} \simeq {\mathrm{400}} \mu{\mathrm m}$. The position resolution
along the beam direction by charge division was less accurate 
($\sigma_{z} \simeq$ 4~cm for protons and deuterons and 10~cm for pions).
The resolution of the mean energy loss, $\sigma(<dE/dx>) / <dE/dx>$, 
was about 15~$\%$ for minimum ionizing particles. The resolution of the
transverse momentum $p_{t}$, 
$\sigma(p_{t})/p_{t}$, was about 4 $\%$ for $p_{t} <$ 0.5 GeV
and worsened to 12~$\%$ near $p_{t} \simeq$ 2~GeV.
The phase space covered by the chamber is shown in Fig.~\ref{acc} 
for the identified $\pi^{\pm}$, protons and deuterons at 1.93~AGeV.
Throughout the paper we use the normalized transverse momentum 
$p_{t}^{(0)} \equiv p_{t} / ( \gamma_{cm} \beta_{cm} m_{0} )$ and 
the normalized rapidity $y^{(0)} \equiv y / y_{cm} - 1$.
Here $y_{cm}$ and $\beta_{cm}$ are the rapidity and velocity of the 
center of mass (c.m.), $\gamma_{cm} = 1 / \sqrt{1 - \beta_{cm}^{2}}$,
and $m_{0}$ is the rest mass of the considered particles
(we use the convention $\hbar$ = c = 1).

\section{particle spectra}

For a more quantitative investigation we present, in Fig.~\ref{bolz}
$\pi^{-}$, proton and deuteron spectra 
in $\Delta y^{(0)}$ bins of 
width 0.1. 
Chosen were two bins at target and at midrapidity
for the reaction at 1.93~AGeV, selected by a cut on the upper end of
the PMUL distribution representing the most central 100~mb of cross section 
(the selectivity of such PMUL cuts on the impact parameter is discussed later). 
The data are plotted in a Boltzmann representation, i.e. 
{\mbox{$1/m_{t}^{2} \cdot d^{2} N/dm_{t} dy^{(0)}$}} vs. $m_{t} - m_{0}$,
in which a thermalized system is expected to show a single-exponential shape
in the absence of flow;
$m_{t} = \sqrt{p_{t}^{2} + m_{0}^{2}}$
and  $m_{0}$ are the transverse mass and the rest mass of the
considered particle, respectively. 
As shown by the dashed lines, the $\pi^{-}$ spectra can be well 
described by the sum of two exponential functions, while the proton and 
deuteron spectra are compatible with one exponential function in our
acceptance (with deviations in the range of very low
$p_{t}$ in the target-rapidity bin):
\begin{equation}
{{1} \over {m_{t}^{2}}} {{d^{2} N} \over {d m_{t} dy^{(0)}}} = 
\left \{
\begin{array}{ll}
C_{l} \cdot e^{- m_{t} / T_{Bl}} + C_{h} \cdot e^{- m_{t} / T_{Bh}}
& \mbox{~~for $\pi^{-}$} \\
C \cdot e^{- m_{t} / T_{B}} & \mbox{~~for p and d.}
\end{array} 
\right.
\label{efit}
\end{equation}
Here, $C$ is a normalization constant, $T_{B}$ is the Boltzmann slope 
parameter, and the subscripts $l$ and $h$ denote the fitting parameters
of the  $\pi^{-}$-spectra in the low- and high-$p_{t}$ range, respectively.
The enhancement of pions at low 
$p_{t}$ can be understood in terms of the $\Delta$(1232) resonance
decay and in fact can be used to determine the number of $\Delta$(1232) at
freeze-out~\cite{bhong1,wein1}. The enhancement 
above the single-exponential fit near zero $p_{t}$ at target rapidity in the 
proton and deuteron spectra which was already
observed at the BEVALAC~\cite{sando1} and at much higher energies at the
AGS~\cite{jd1} is attributed to spectator contributions.
That is why it is seen only at target rapidity; it 
disappears as soon as the window is moved 
by a step of only 0.2 in $y^{(0)}$ towards midrapidity. 
To obtain the rapidity distributions $dN / dy^{(0)}$, we integrated 
the fitted functions of Eq.~(1) from $p_{t} =$ 0 to $\infty$ 
in order to account for the missing part in the acceptance
(cf. Fig.~\ref{acc}). 
Both because of these limitations and the fact that we did not try to
include the mentioned low-$m_{t}$ enhancement of protons and deuterons
near target rapidity (Fig.~\ref{bolz}) the resultant rapidity distributions are
expected to represent primarily the distribution of {\em participant}
matter to which our prime interest is addressed.

In the spectra presented in this paper only the statistical errors
are shown. The systematic errors in the $dN / dy^{(0)}$ distributions 
are different for each particle species and vary with the beam energy
and rapidity, the largest values existing for the midrapidity
data at 1.93 AGeV, mainly due to the smallest geometrical acceptance.
At this energy the integration over the complete $p_{t}$
range leads to estimated errors of about
2, 5 and 20 $\%$ for pions, protons and deuterons at midrapidity,
respectively. These values were determined by
using three different possibilities, i.e. an exponential shape in
Boltzmann and invariant representations and the thermal model including
the radial flow (Eq.~\ref{shellflow} in Sec. V).
The uncertainty in the tracking efficiency (10 $\%$) was obtained 
by visual inspection of several hundred event displays and by comparing 
the results of an analysis with different tracking algorithms~\cite{pelte1}. 
The uncertainty in the particle identification (2 $\%$) was 
determined by changing the software criteria. 
Assuming that the sources of these different errors add incoherently, 
we obtain maximal systematic errors for the $dN / dy^{(0)}$ distributions 
of 10, 11 and 23 $\%$ for pions, protons and deuterons, respectively.
A confirmation that the extrapolation procedure over the full $p_{t}$-range 
is reasonable is the integrated charge of all reaction 
products up to $^{4}$He which for the most cental collisions (100 mb) agrees
within 10 $\%$ with the total charge of the system no matter
which spectral form is assumed. The systematic error in  $T_{B}$,
caused by the different tracking methods and the 
different fitting range in $p_{t}$ is 
estimated to about 5 $\%$.

\subsection{Centrality Dependence}

The centrality dependence was studied in the case of the reaction
at 1.93 AGeV. 
Figs.~\ref{c_slope} and \ref{c_dndy} show the $T_{B}$ and 
$dN / dy^{(0)}$ distributions for different cuts on PMUL thereby varying
the selected cross section between the most 
central 100~mb and 420~mb, representing 4 and 15~$\%$ of the total reaction 
cross section, respectively. The relation between the centrality cuts 
and the impact parameter $b$ as calculated within 
the IQMD model are summarized in Table~\ref{Tcent}. Note that the data 
are measured only for $y^{(0)} <$ 0, and then reflected about 
midrapidity, using the symmetry of the colliding system.
It is apparent that the $T_{B}$ distributions of Fig.~\ref{c_slope}
are practically identical for 
the three centrality cuts both for protons and for each of the two pion 
components, whereas in Fig.~\ref{c_dndy}
a relatively modest enhancement of the midrapidity yield
is observed with increasing
centrality in the $dN / dy^{(0)}$ distributions of both particles, somewhat
more pronounced in case of the pions.
When cutting even sharper on centrality we find
a saturation of the pion value at about 30~mb, 
while the proton distribution remains stable below 100 mb.
The lack of dramatic changes below an effective sharp-cut impact parameter
of about 4 fm indicates a limited impact parameter resolution when using
the PMUL selection in a relatively small system such as Ni+Ni.

In the case of global thermal equilibrium one expects the following 
characteristic rapidity dependence of the slope parameter $T_{B}$
\begin{equation}
T_{B}(y) = T/\cosh (y)
\label{coshy}  
\end{equation}
leading to a bell-shaped curve for $T_{B}$ with $T_{B}=T$ at midrapidity
($y=0$).
This dependence, adjusted to the experimental $T_{B}$ (125 MeV for protons)
at midrapidity, is compared to the experimental results in the lower panel
of Fig.~\ref{c_slope}.
It is evident that the data are not described by this global equilibrium
assumption, but rather imply that baryonic matter is
'colder' away from midrapidity.
The failure of the purely thermal scenario is also evident from the experimental
proton $dN/dy^{(0)}$ distributions (lower panel Fig.~\ref{c_dndy}):
they are much wider than the calculated one for an isotropically emitting
thermal source of 125 MeV temperature.

In contrast to the protons we find that the $T_{B}$ and $dN/dy^{(0)}$
distributions of the  high $m_{t}$ component of the pions {\em are}
compatible with an apparent global temperature which is however somewhat
lower ($T_{\pi }=115$ MeV) than implied by the midrapidity value of $T_{B}$ for
the protons (see Figs.~\ref{c_slope} and \ref{c_dndy}).
When including the deuteron data (Fig.~\ref{e_slope} and \ref{e_dndy}) into the 
considerations one notices at midrapidity a pronounced increase of
$T_{B}$ with the mass of the particle.

The fact that the {\em full} rapidity distributions of protons indicate
incomplete thermalization, while the pion rapidity distributions appear to
be 'thermal' could indicate any or a combination of both of the following
possibilities:
only a (midrapidity) fraction of the nucleons are part of a thermally
equilibrated fireball that comprises most of the pions {\em or}
alternatively, there is no fully equilibrated fireball, as suggested by the
nucleonic distributions, while the $dN/dy$ distributions of the rather
light pions are insensitive to this nonthermal behaviour.
Exploring the first possibility, but not definitely excluding the second 
we shall see later that it is possible to describe the 
{\em midrapidity} $m_{t}$
spectra of protons, deuterons and pions                               
with a common temperature if one introduces {\em flow}
(see section V).

\subsection{Beam Energy Dependence}

Figs.~\ref{e_slope} and \ref{e_dndy} show the beam energy dependence 
of $T_{B}$ and $dN / dy^{(0)}$ distributions of  $\pi^{-}$,
protons and deuterons under a PMUL cut of 100~mb. Generally one finds
larger $T_{B}$ values, especially near midrapidity, when going 
from 1.06~AGeV to 1.93~AGeV.
The high-$p_{t}$ slope parameter $T_{Bh}$ of the $\pi^{-}$ spectra changes more
than the low-$p_{t}$ slope parameter $T_{Bl}$ (an increase at midrapidity
of 24~$\%$ compared to
only 11 $\%$). The increase 
becomes larger for the heavier particles (30 $\%$ for protons and 34 $\%$ 
for deuterons), which hints to a larger flow velocity at the higher beam
energies.
Remarkably there is no significant change of $T_{B}$ with beam energy near
target (projectile) rapidity.
In case of the rapidity distributions $dN / dy^{(0)}$ of
Fig.~\ref{e_dndy}, we find that the spectra of protons 
and deuterons exhibit similar shapes at the different energies. This implies 
that the width of the baryon $dN / dy^{(0)}$ spectra is rather independent 
of the beam energy, which will be addressed further in the next section.
The inverse slope parameters at midrapidity and the integrated
particle yields per event are summarized in Table~\ref{Tnumbers}.
For a more thorough discussion of the pion yields and the systematic errors
affecting them we refer to~\cite{pelte1,dp1}.

\section{baryon rapidity spectra}

Baryon rapidity distributions allow a view on the stopping 
power of nuclear matter, provided the uncertainties introduced 
by the limited knowledge of the collision geometry
can be controlled.
Especially for the present light system, finite particle number
fluctuations are important when trying to select head-on collisions.
In order to understand the properties of our selection
criterion PMUL, 
we first compare the proton and deuteron 
$dN / dy^{(0)}$ spectra with the IQMD model~\cite{bass1}.

\subsection{Comparison with IQMD}

As an example we compare in Fig.~\ref{dndyiqmd},
the experimental $dN / dy^{(0)}$ distributions of protons and deuterons
 at 1.93~AGeV for the most central PMUL cut (100 mb) to IQMD model results,
obtained with the option of a hard EoS (compression constant $K =$ 380 MeV)
and a momentum dependent potential, IQMD(HM). The  
solid lines represent the model results under the same centrality cut as the 
data, i.e. a PMUL cut on the upper 100~mb in 
the charged-particle multiplicity spectra for 
${\mathrm{7}}^{\circ} \le \theta_{L} \le {\mathrm{30}}^{\circ}$.
The dashed lines represent results obtained with the
corresponding impact parameter cut. 
The difference between the solid and dashed lines 
shows the uncertainty imposed by using PMUL as a centrality criterion 
instead of the exact $b$. It turns out that the effect is negligible for 
protons, and small
for deuterons.
Composite particles from the model were formed by a space coordinate
cluster algorithm after an elapsed collision time of 200 fm/c using the
standard distance parameter of 3 fm.
Under these conditions IQMD underpredicts the deuteron-to-proton ratio by
approximately a factor of five.
Since we wish to emphasize the shapes of the rapidity distributions all
model calculations are normalised to the integral of the data in the
figure.
We conclude that IQMD reproduces the degree of stopping rather well for the
most central collisions. The small 'spectator' shoulder for protons 
in the model (or the slight peak in the case 
of the deuterons) is not seen in the data because, as mentioned earlier,
the spectator components of the proton and deuteron spectra 
at low-$p_{t}$ are suppressed by our integration.
The comparison of measured and simulated $dN / dy^{(0)}$ spectra of
protons and deuterons in case of the other
beam energies shows a similar degree of agreement for the 100~mb cut.
Therefore, we conclude that the IQMD model
reproduces the shape of the measured proton
$dN / dy^{(0)}$ spectra at the most central collisions.
IQMD calculations with a soft EoS ($K =$ 200 MeV)
and momentum dependent potential, IQMD(SM),  
show very similar results as the IQMD(HM) version, so there is 
no sensitivity
to the stiffness of the EoS in the proton and deuteron rapidity spectra.

Based on these observations, we compare in Fig.~\ref{dndyiqmdauni}
the $dN / dy^{(0)}$ model predictions for protons 
and deuterons in Ni + Ni collisions
(right panel) calculated for zero impact parameter in 
the IQMD model with results of the isotropic expansion 
model, using the parameters given in Sec. V (Table~\ref{Tflow}).
Also shown are IQMD predictions for  
Au + Au collisions of the same energy of 1.06 AGeV (left panel) 
together with the result of an isotropic expansion scenario using the
parameters $T =$ 81 MeV and $\beta_{r} =$ 0.32 at 1.0 AGeV from  
Ref.~\cite{mal1}. In the IQMD model the widths of the $dN / dy^{(0)}$ 
distributions of protons and deuterons in Au + Au are narrower 
than in Ni + Ni collisions at the same beam energy. Additionally, the 
$dN / dy^{(0)}$ for Au + Au collisions can be described nicely by the 
isotropic expansion model, while the one for Ni + Ni is wider 
than the model, in accordance with our experimental
findings (cf. Fig.~\ref{c_dndy}).

In principle, one can not distinguish incomplete stopping from a 
longitudinal expansion after full stopping on the basis of
the rapidity spectra alone. However,
the systematic comparison of the $dN / dy^{(0)}$ spectra between the small 
and large colliding system within the IQMD model, the results of
which are supported by our experimental data
in the case of the most central Ni + Ni collisions, can 
help to resolve this ambiguity. The narrower $dN / dy^{(0)}$ shape
of Au + Au as compared to Ni + Ni for $b =$ 0 fm indeed tells us that the
IQMD model predicts a partial transparency for the latter system.
One would expect a wider $dN / dy^{(0)}$ distribution or a smaller mean 
rapidity shift $\delta y_{p}$ (as defined in the next chapter) 
for heavier colliding systems in case of a 
longitudinal expansion after full stopping. Using the same model,
this subject was investigated by Bass $et~al.$ by means of the
(n-p)/(n+p) ratio in the isospin-asymmetric system  $^{50}$Cr + $^{48}$Ca,
where a partial transparency was also predicted at an energy of 1~AGeV 
\cite{bass2}.

\subsection{Scaling of the Mean Rapidity Shift of Protons}

Recently, Videb$\ae$k and Hansen discussed the systematics 
of the baryon rapidity losses in central nucleus-nucleus collisions 
at AGS and SPS energies~\cite{fv1}. The main conclusion was that the 
mean rapidity losses scaled with the beam rapidity from 10 to 
200 AGeV. In this section we want to study whether this scaling behavior
holds at the present lower beam energies, too, i.e. down to 1 AGeV. 

Table~\ref{Tyshift} summarizes the results of our analysis and the one of
Ref.~\cite{fv1} 
in terms of the mean rapidity 
shift of protons ($\delta y_{p}$) defined as
\begin{equation}
\delta y_{p} \equiv {{\int_{-\infty(y_{cm})}^{y_{cm}(\infty)} 
| y - y_{t(b)} | ~(dN_{p}/dy)  ~dy} \over 
{\int_{-\infty(y_{cm})}^{y_{cm}(\infty)}~(dN_{p}/dy) ~dy}},
\label{yshift}
\end{equation}
where $y_{t}$ and $y_{b}$ represent the target and beam rapidities, 
respectively, and $dN_{p}/dy$ is the proton rapidity distribution.
The quantity $\delta y_{p}$ reflects the inverse width of the rapidity
distribution: The more protons, or baryons in general, pile up at the
c.m.-rapidity, the higher are the $\delta y_{p}$ values. 
The scaled shift $\delta y_{p} / y_{b}$
is shown in Fig.~\ref{dyproton} as a function of beam energy.
For the relatively smaller systems (Ni + Ni at the SIS, Si + Al at 
the AGS and S + S at the SPS) $\delta y_{p} / y_{b}$ is constant 
for the beam energies between 1 and 200 AGeV, which implies 
that the shape of the baryon $dN / dy^{(0)}$ spectra is independent 
of the beam energies over this energy range. For the heavier system
(Au + Au) both the IQMD prediction at 1 AGeV and the data at 11 AGeV show
a slightly larger $\delta y_{p}/y_{b}$, which means a higher concentration of 
baryons at midrapidity. 

\section{Radial Flow}

There has been a lot of effort to understand the collective motion in
heavy ion collisions, hoping to get a handle on the nuclear equation of 
state~\cite{hsto1}. Especially the radial flow of the midrapidity
fireball as an important energy 
carrier~\cite{wr1,mal1,mpt1,scj1,pbm1,poggi1}
has been studied extensively. In this section we want to
 extract the temperature $T$ and the average radial 
flow velocity $\beta_{r}$ from the midrapidity transverse-momentum 
spectra. We employ the formula of the simple thermal blast model proposed by 
Siemens and Rasmussen~\cite{ps1}:
\begin{equation}
{{1} \over {m_{t}^{2}}} {{d^{2} N} \over {d m_{t} dy^{(0)}}} \propto
\cosh y \cdot e^{-\gamma_{r} E / T} \cdot 
[(\gamma_{r} + {{T}\over{E}}){{\sinh \alpha} \over {\alpha}} -
{{T} \over {E}}\cosh \alpha],
\label{shellflow}
\end{equation}
with $\gamma_{r} = 1 / \sqrt{1-\beta_{r}^{2}}$ and 
$\alpha = (\gamma_{r} \cdot \beta_{r} \cdot p) / T$, where
$E = m_{t} \cosh y$ and $p = \sqrt{p_{t}^{2} + m_{t}^{2} \sinh^{2} y}$ 
are the total energy and momentum of the particle in the c.m. system. 
In this model, the thermally equilibrated system
expands isotropically, 
then freezes out suddenly at which time
all the particles in the system share
a common local $T$ and $\beta_{r}$.

We are aware that the full event topology is {\em not} isotropic and the
Ansatz Eq. (4) can therefore at best describe a part of the populated phase
space which we restrict to the midrapidity interval 
 ($-0.1 < y^{(0)} < 0.0$)
under the 100 mb PMUL cut. This should
minimize the contaminations by spectators and non-isotropic flow 
components (Fig.~\ref{c_dndy}). 
The effect of the collective flow can be more significant for heavier 
particles as seen at lower beam energy \cite{wr1,poggi1}, 
but at the present energies composite particles are so few that we
restrict the analysis to pions, protons and deuterons. 
For such light particles it was shown in Ref.~\cite{wr1} that details of
the flow profile are not discernable because the thermal fluctuations wash
them out. 
This justifies the use of the Siemens-Rasmussen formula which replaces the
integration over a complex flow velocity profile by a single
'representative' velocity $\beta_{r}$.
On the other hand,
the contamination of the proton and deuteron spectra
by products evaporated  
from the heavier fragments is largely reduced in the energy range 
we are studying here.
Besides simplicity, an important benefit of this simple-minded fit to the
data is that a direct comparison with the results of a very similar
analysis for the Au + Au system \cite{mal1} can be done.

Applying the definition of a slope parameter in Eq.~\ref{efit} to 
Eq.~\ref{shellflow}, we evaluate an effective slope $T_{B}^{eff}$ 
from the model as follows
\begin{equation}
{T_{B}^{eff}} \equiv - [{{d} \over {d m_{t}}} 
\{\ln({{1} \over  {m_{t}^{2}}}{{d^{2} N} \over {dm_{t}dy^{(0)}}})\}]^{-1}.
\label{tbcalc}
\end{equation}
Here $T_{B}^{eff}$ shows a combined effect of $T$ and $\beta_{r}$, and
the model gives the estimate of $T_{B}^{eff}$ at each $m_{t}$ value. 
In the top panel of Fig.~\ref{auflow}, our data are shown by 
bold lines (the fitting errors are smaller than the thickness of each line)
together with the model calculations. The $m_{t}$ range of the fit to the 
experimental data, which is another important constraint in determining the 
model parameters, is also indicated. We include only $T_{Bh}$ of $\pi^{-}$
since the low $p_{t}$ component of the pion spectra is strongly 
affected by the $\Delta$(1232) decay~\cite{bhong1,wein1}. 
To determine $T$ and $\beta_{r}$, the two parameters were varied until
the model describes our experimental data;
the resulting values are summarized in Table~\ref{Tflow}.  
The top panel of Fig.~\ref{auflow} displays 
the range of the model calculations for $\pi^{-}$ (horizontally hatched), 
protons (vertically hatched) and deuterons (diagonally hatched). 
Having determined $T$ and $\beta_{r}$, we confirm the results by comparing 
the spectra from the model with the data directly as shown
in the bottom panel of Fig.~\ref{auflow}. The results were also checked 
by the simultaneous fitting method requiring a minimum $\chi^{2}$ per
degree of freedom, and they are consistent with each other within 5 $\%$.

Before discussing the results in the framework of general flow systematics
it is worthwhile to check on two points:\\
1) could the fact that resonances other than the $\Delta(1232)$ are excited
(but not explicitly treated in the analysis of the high momentum part of
the pion spectra) strongly modify the analysis and \\
2) in view of the fact that the spectral shape analysis requires only
(local) {\em thermal} equilibrium, can we check that the particle {\em
yields} are consistent with {\em chemical} equilibrium?

We have studied both questions 
in the context of a hadrochemical equilibrium model~\cite{pbm1,bhong2}.
The model parameters, chemical freeze-out temperature $T_{C}$ 
and baryon chemical potential $\mu_{B}$, 
are treated as free parameters, and fixed to reproduce 
the experimental yields of nucleons, deuterons, thermal pions, 
$\Delta$(1232)-~\cite{bhong1} and $\eta$-mesons
from N$^*$(1535) resonances~\cite{taps1}.  
The extracted parameters, $T_{C}$ and $\mu_{B}$, are also shown in
Table~\ref{Tflow} (for a comparison of the model results with experimental
particle yield ratios, see~\cite{bhong1}).
At all energies the temperature $T$ derived from the spectra including flow
agrees with the chemical freeze-out temperature
$T_{C}$ obtained from the particle yields within 
8~$\%$~\cite{bhong2}.
The extracted baryon chemical potentials correspond to roughly  $0.5\pm0.2$
of the saturation density (0.17 fm$^{-3}$).
 
The effects due to higher resonances were estimated using fixed average
masses.
Within this model, the total freeze-out population
of N$^{\star}$(1440), N$^{\star}$(1520) and N$^{\star}$(1535) is estimated 
to be less than 10 $\%$ of the $\Delta$(1232) population at 1.93 AGeV.
For freeze-out densities of approximately half the normal nuclear matter
density, the contribution of thermal pions compared to the number of pions
from resonances with masses larger than the $\Delta (1232)$ resonance
exceeds the latter by more than a factor 10.
Therefore it is reasonable to assume in the analysis
that the high-$p_{t}$ component of 
pion spectra is due to the thermal pions.

Fig.~\ref{tandb} shows the comparison of the model parameters 
$T$ and $\beta_{r}$ and the fraction of the total available energy per nucleon 
in c.m. contained in the radial flow motion 
($E_{rflow} / E_{cm}$, with $E_{rflow} = (\gamma_{r} - 1) m_{N}$ 
and $m_{N}$ being the nucleon mass) from the current analysis with other 
results for the system Au + Au~\cite{wr1,mal1,poggi1}.
Before drawing conclusions from the data presented in Fig.~\ref{tandb} the
reader should be aware of the specific differences in the experimental
analyses.
While the present analysis used midrapidity pion (high T component), proton
and deuteron spectra, the EOS collaboration~\cite{mal1} based their
conclusions primarily on the $90^{\circ}$ (c.m.) spectra of A=2,3 and 4
fragments, but otherwise the same formalism was used as described here.
The flow analysis of Ref.~\cite{poggi1} was based essentially on a
comparison of the mass dependence of the average kinetic energies of Z=1
fragments (i.e. p,d,t). The temperatures were derived in a more indirect
way with the help of simulations taking into account evaporation; for the
values cited in the figure a freeze-out
density of 80\% of the ground state density was used which gave the best overall
reproduction of the spectral shapes of all the Z=1 and 2 isotopes.
Concerning the flow velocities of Ref.~\cite{poggi1} we have added a
Coulomb correction to the published values.
The analysis of \cite{wr1} extended to fragments with Z=2-8 and to the full
measured phase space allowing for a more complex flow profile under the
additional constraint of energy conservation.

While some of the straggling in the data points shown in Fig.~\ref{tandb}
could well be due to the different methods and data types used to extract
the parameters, one can nevertheless discern some general trends.
First of all we find that both $T$ and $\beta_{r}$ increase monotonically
as a function of beam energy up to 2 AGeV for both colliding systems. 
Secondly the temperature $T$ seems to be independent of the system, 
while the radial flow velocity $\beta_{r}$ is larger for the larger system 
size (at least close to 1~AGeV ). Intuitively this system size dependence 
of $\beta_{r}$ is consistent with our conclusion of the nuclear stopping 
power in the previous section. The larger nuclei have a larger stopping 
power, accumulate more pressure in the midrapidity fireball, 
and as a result the system expands faster.
The fact that the 'freeze-out' temperature turns out to be the same for the
Ni as for the Au system does not necessarily mean that the 'primordial'
temperature, prior to expansion, was the same.
As a matter of fact, within the picture of an adiabatic expansion following
maximal compression, one is tempted to conclude that the primordial
temperature was somewhat higher in the Au system since the larger
collective energy at freeze-out in the heavier system indicates a stronger
conversion of thermal energy, hence a stronger cooling.
If this tentative interpretation is true, then 'participant' matter in the
Ni system, even around midrapidity, has not thermalized as thoroughly as in
the Au system at the stage of maximal compression.
This shows the necessity to study system size dependences in order to
assess quantitatively the degree of equilibration in such collisions.

As a third point we wish to emphasize that the flow energy represents a
sizeable fraction of the available energy $E_{cm}$ (lower panel of
Fig.~\ref{tandb}).
In the context of the present work we note that the flow energy in the Ni
system (13\% of $E_{cm}$ as mean value of the three Ni points)
is about half as large as that deduced for the Au
system at a comparable energy.
From the yields and temperatures (Tables II and IV) we estimate that the
baryons take up 50\% (at 1.06A GeV), resp. 35\% (at 1.93A GeV) of $E_{cm}$
as thermal energy, pion production consumes about 10\% (at 1.06A GeV), resp.
17\% (at 1.93A GeV).
A simple energy balance consideration then leads to the conclusion
 that a significant 
fraction (approximately 30 $\%$) of $E_{cm}$ is left in surplus
longitudinal movement of the leading baryons for relatively small colliding 
systems such as Ni + Ni. One should, however,
not argue too strongly about the precision of these numbers: the different 
analysis methods imposing the energy conservation, freeze-out density 
and flow velocity profiles may favor different sets of $T$ and $\beta_{r}$. 

\section{conclusions}

We have studied in detail the $\pi^{-}$, proton and deuteron
spectra for 
central Ni + Ni collisions at beam energies between 1 and 2 AGeV.
We do not observe any 
dependence of the slope parameter of the transverse momentum spectra 
on the event centrality (from 420 mb to 100 mb), 
while a higher pion production and a stronger proton concentration at
midrapidity is seen for more central events. 

The slope parameters are generally larger for higher beam energies, and 
the effect is more pronounced for the heavier particles; 
the pion slope parameter for the high 
transverse-momentum component changes more than the one of the 
low transverse-momentum component. The rapidity spectra of protons and 
deuterons show very similar shapes at the different bombarding energies
under the same centrality cuts.
These shapes are however incompatible with global thermal equilibrium.

The IQMD model can reproduce the measured proton and deuteron rapidity 
spectra for the most central events, with very similar results 
for the option of a hard and soft equation of state. The rapidity spectra 
of protons and deuterons from the IQMD in Au + Au are narrower  
than in Ni + Ni for vanishing impact parameter, which implies more 
nuclear stopping power in larger colliding system. 
The known scaling law for the mean rapidity shift of 
protons (scaled with the beam rapidity) is satisfied down to 1 AGeV 
for the small colliding system size.

The freeze-out temperature and radial flow velocity
of midrapidity particles increase
with the beam energy up to 2 AGeV, but only the radial flow velocity shows
a system size dependence. The energy fraction consumed by the 
radial flow motion is constant for a given colliding system size at 
beam energies between 0.1 and 2 AGeV. In Ni + Ni collisions
this energy fraction is about one half of the value found 
in Au + Au collisions.

\acknowledgments 
We would like to thank Prof. Peter Braun-Munzinger for many discussions
and proofreading of the manuscript.
This work was supported in part by the Bundesministerium
f\"{u}r Forschung und Technologie under the contracts 06 HD 525 I(3),
06 DD 666 I(3), X051.25/RUM-005-95 and X081.25/N-119-95.

\begin{table}
\caption{ 
Centrality cuts on the PMUL distributions, related cross sections ${\sigma}$
and the corresponding impact 
parameters $b$, calculated with 
the IQMD model with hard equation of state (EoS) and a  
momentum dependent potential.}
The nuclear radius $R(A)$ is given by 
1.2$A^{1/3}$, where $A$ is the number of nucleons; the maximum impact 
parameter $b_{max}$ is determined by $\sqrt{{\sigma} / {\pi}}$.

\vspace{0.5cm}
\begin{tabular}{cccccc} 
System & $R(A)$(fm) &  PMUL & $\sigma$(mb) & $b_{max}$(fm) & 
$<b>_{IQMD}$(fm) \\ \hline
                  &     &  $\geq$25 & 420 & 3.7 & 2.6  \\
1.93 AGeV Ni + Ni & 4.7 &  $\geq$31 & 250 & 2.8 & 2.1  \\
                  &     &  $\geq$37 & 100 & 1.8 & 1.6  \\ \hline
1.06 AGeV Au + Au & 7.0 &  $\geq$76 & 120 & 2.0 & 2.8  \\
\end{tabular}
\label{Tcent}
\end{table}
\vspace{1.0cm}

\begin{table}
\caption{
Inverse slope parameters at midrapidity $T_{B}^{0}$
and integrated particle yields per event 
with a cut on PMUL of 100 mb (cf. Table~I).
For pions the two slopes with index $l$ and $h$ refer to the low- and
high-$p_{t}$ part of the spectra.
Only the dominant systematic 
errors are quoted.
The statistical errors are about 20 and 10 $\%$ of the systematic errors
of $T_{B}^{0}$ and the yields, respectively.
}
\vspace{0.5cm}
\begin{tabular}{ccccccc}
 & \multicolumn{2}{c}{$\pi^{-}$} & \multicolumn{2}{c}{proton} & 
   \multicolumn{2}{c}{deuteron} \\ \cline{2-3} \cline{4-5} \cline{6-7}
$E_{beam}/A \mathrm{(GeV)}$ & $T_{Bl}^{0}/T_{Bh}^{0}$ (MeV) & Yield &
$T_{B}^{0}$ (MeV) & Yield & $T_{B}^{0}$ (MeV) & Yield \\ \hline
1.06 & 55$\pm$3/93$\pm$5 & 3.6$\pm$0.4 & 96$\pm$5 & 38.4$\pm$4.2 &
104$\pm$5 & 14.4$\pm$2.4 \\
1.45 & 56$\pm$3/100$\pm$5 & 5.8$\pm$0.6 & 111$\pm$6 & 41.6$\pm$4.6 &
120$\pm$6 & 12.8$\pm$2.6 \\
1.93 & 61$\pm$3/115$\pm$6 & 8.5$\pm$0.9 & 125$\pm$6 & 44.0$\pm$4.8 &
139$\pm$7 & 11.6$\pm$2.4 \\
\end{tabular}
\label{Tnumbers}
\end{table}
\vspace{1.0cm}

\begin{table}
\caption{
Summary of the mean rapidity shift of protons (see text for definition).
Note that the present analysis 
includes protons and deuterons. The numbers in parenthesis are the 
results from the IQMD model calculations with a hard EoS.
}
\vspace{0.5cm}
\begin{tabular}{ccccccc} 
$E_{beam}/A \mathrm{(GeV)}$ & $y_{b}$ & System & $\sigma / \sigma_{tot}(\%)$ 
 & $\delta y_{p}$ & $\delta y_{p} / y_{b}$ & Reference \\ \hline
1.06 & 1.388 & Ni+Ni & 3.6 & 0.389 & 0.280 & this work \\
 & & & ($b =$ 0 fm) & (0.394) & (0.284) & IQMD(HM) \\
1.06 & 1.388 & Au+Au & ($b \le$ 0.5 fm) & (0.432) & (0.311) & IQMD(HM) \\
1.45 & 1.586 & Ni+Ni & 3.6 & 0.453 & 0.285 & this work \\
1.93 & 1.782 & Ni+Ni & 3.6 & 0.503 & 0.282 & this work \\
 & & & ($b =$ 0 fm) & (0.521) & (0.292) & IQMD(HM) \\
11.6 & 3.21 & Au+Au & 4.0 & 1.02 & 0.32 & Ref. 21 \\ 
14.6 & 3.44 & Si+Al & 7.0 & 0.97 & 0.28 & Ref. 21 \\ 
200. & 6.06 &  S+S  & 3.0 & 1.69 & 0.28 & Ref. 21 \\ 
\end{tabular}
\label{Tyshift}
\end{table}
\vspace{1.0cm}

\begin{table}
\caption{
Summary of radial flow velocities $\beta_{r}$ 
and  freeze-out temperatures $T$ 
for the high-$p_{t}$ part of the pion spectra and
of the proton and deuteron distributions
in Ni + Ni collisions for the studied energies,
derived within the model of Siemens and Rasmussen~[22]. 
The parameters  $T_{C}$ and $\mu_{B}$ from the chemical equilibrium 
model~[34] are also included. 
}
\vspace{0.5cm}
\begin{tabular}{ccccc} 
$E_{beam}/A$(GeV) & $\beta_{r}$ & $T$ (MeV) & $T_{C}$ (MeV) & $\mu_B$\\ \hline
1.06 &  0.23 $\pm$ 0.03 & 79 $\pm$ 10 & 73 $\pm$ 10 & 780 $\pm$ 30\\
1.45 &  0.29 $\pm$ 0.03 & 84 $\pm$ 10 & 81 $\pm$ 10 & 755 $\pm$ 25\\
1.93 &  0.32 $\pm$ 0.04 & 92 $\pm$ 12 & 90 $\pm$ 13 & 725 $\pm$ 35 \\
\end{tabular}
\label{Tflow}
\end{table}
\vspace{1.0cm}

\begin{figure}
\caption{ 
Acceptance of $\pi^{\pm}$, protons and deuterons at 1.93 AGeV under a 
centrality cut of 420~mb on PMUL (see text).
With the definition of $y^{(0)}$, -1, +1 and 0 denote the 
target-, projectile- and the c.m.-rapidity. Each successive contour line 
represents a relative factor of two in terms of yields. The dash-dotted 
lines show the geometrical limit of the drift chamber at 
$\theta_{L} = {\mathrm{30}}^{\circ}$, while the dotted line for 
$\pi^{+}$ shows the high $p_{Lab}$-cut imposed by the separation against
the protons.
}
\label{acc}
\end{figure}

\begin{figure}
\caption{ 
Boltzmann spectra (absolute yield per event) of $\pi^{-}$ 
(solid circles), protons (open circles) and deuterons (solid triangles) at 
target rapidity ($y^{(0)}$ from -1.0 to -0.9, left) and at midrapidity
($y^{(0)}$ from -0.1 to 0.0, right).
The data are for 1.93 AGeV collisions with a cut of 100~mb on PMUL.
The $\pi^{-}$ specta are multiplied by 100 for a clearer display.
The dashed lines show the fits with the sum of two exponentials for pions and
with one exponential for protons and deuterons.
}
\label{bolz}
\end{figure}

\begin{figure}
\caption{ 
Centrality dependence of the Boltzmann slope parameter $T_{B}$ 
for $\pi^{-}$ (separately for the low- and high-$p_{t}$ component)
and protons at 1.93 AGeV.
The solid lines are the results of the isotropic expansion 
model without collective radial flow. 
For details see Sec. V.
}
\label{c_slope}
\end{figure}

\begin{figure}
\caption{ 
Centrality dependence of the $dN / dy^{(0)}$ spectra for $\pi^{-}$ 
and protons at 1.93 AGeV. The solid lines are the results of the 
isotropic expansion model without collective 
radial flow. For details see Sec. V.
}
\label{c_dndy}
\end{figure}

\begin{figure}
\caption{ 
Beam energy dependence of the Boltzmann slope parameter $T_{B}$ for
$\pi^{-}$ (low- and high-$p_{t}$ component), 
protons and deuterons under a cut of 100~mb on PMUL. 
}
\label{e_slope}
\end{figure}

\begin{figure}
\caption{ 
Beam energy dependence of the $dN / dy^{(0)}$ spectra for
$\pi^{-}$, protons and deuterons with a cut of 100~mb on PMUL. 
For the deuteron data the error bars denote
the systematic errors due to the different 
extrapolations towards $p_{t} =$ 0. 
}
\label{e_dndy}
\end{figure}

\begin{figure}
\caption{ 
Comparison of experimental proton (left) and deuteron (right)
$dN / dy^{(0)}$ spectra (symbols) with results of the
IQMD(HM) model at 1.93 AGeV (solid and dashed lines). The solid IQMD curves were
obtained under the same cut on PMUL as in the data analysis, 
the dashed lines under the corresponding impact parameter cut.
}
\label{dndyiqmd}
\end{figure}

\begin{figure}
\caption{
Comparison of $dN / dy^{(0)}$ spectra from the IQMD(HM) model (symbols)
with results of the isotropic expansion model (solid lines) for the systems
Au + Au (left) and Ni + Ni (right), both at 1.06 AGeV.
The IQMD results are for zero impact
parameter ($b \le$ 0.5 fm for Au + Au, $b = 0$ for Ni + Ni).
The isotropic expansion model
uses the parameters given in Table III for Ni + Ni and those of 
Ref.~[11] for Au + Au.
}
\label{dndyiqmdauni}
\end{figure}

\begin{figure}
\caption{ 
Mean rapidity shift of protons scaled with the beam rapidity
as a function of beam energy.
The dashed straight line at 0.28 is only to 
guide the eye. The data of the AGS and SPS experiments are from
Ref.~[21].}

\label{dyproton}
\end{figure}

\begin{figure}
\caption{ 
Top: $T_{B}^{eff}$ vs. $m_{t}$ of calculations (see text for details)
within the isotropic expansion model spanned by 
the given parameters of $T$ and $\beta_{r}$ for $\pi^{-}$ (horizontally 
hatched), proton (vertically hatched) and deuteron (diagonally hatched)
at three beam energies.
Bottom: Measured Boltzmann spectra compared to the model calculations 
(solid lines). Note that the $\pi^{-}$ spectrum is multiplied by 100
for a clearer display.
}
\label{auflow}
\end{figure}

\begin{figure}
\caption{ 
Compilation of results of various experiments, showing
$T$, $\beta_{r}$ and $E_{rflow} / E_{cm}$, 
the fraction of available c.m. energy contained in radial flow,
as function of the beam energy. 
}
\label{tandb}
\end{figure}

\setcounter{figure}{0}
\pagebreak
\begin{figure}[hhh]
.
\includegraphics{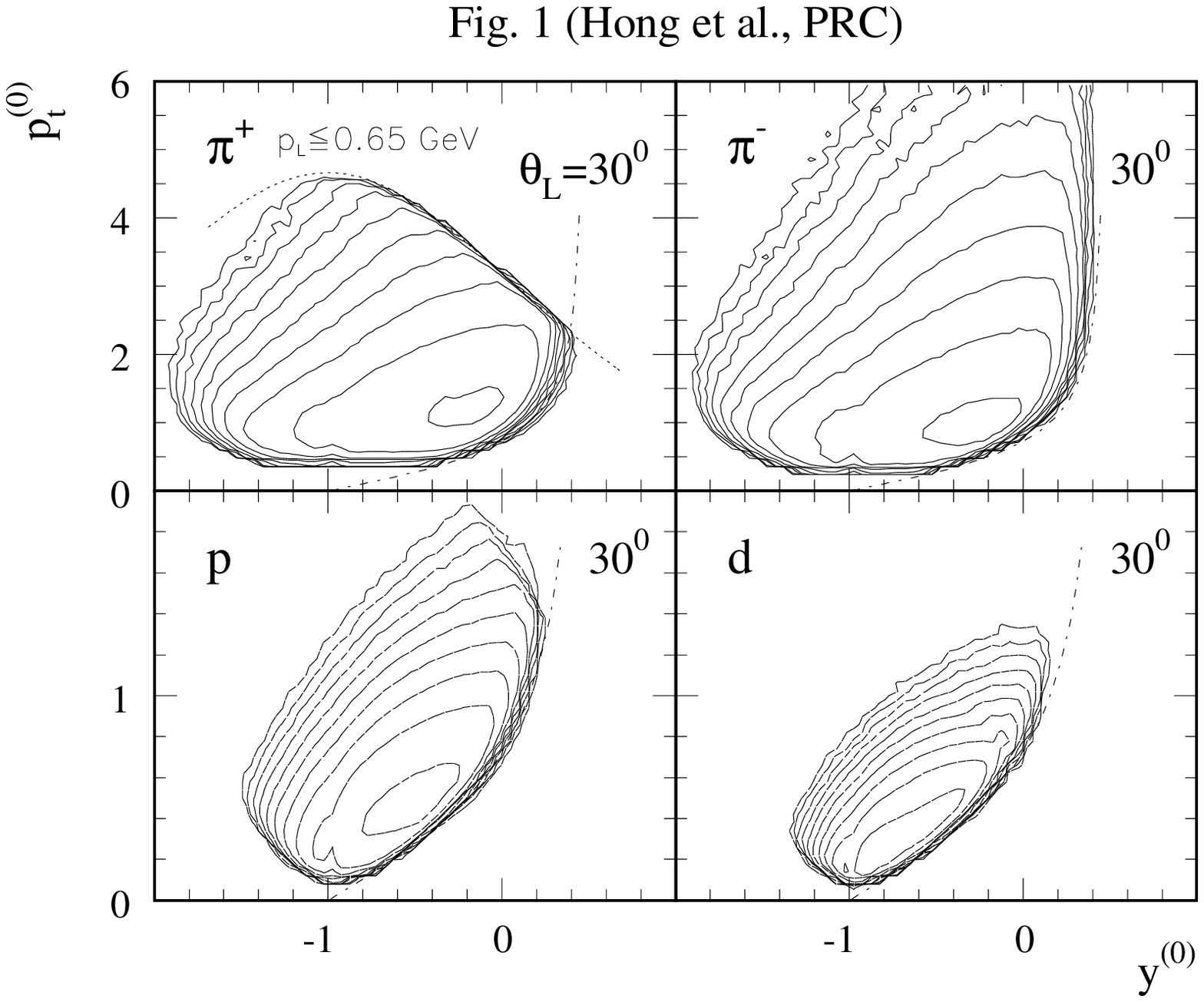}
\end{figure}

\pagebreak
\begin{figure}[hhh]
.
\includegraphics{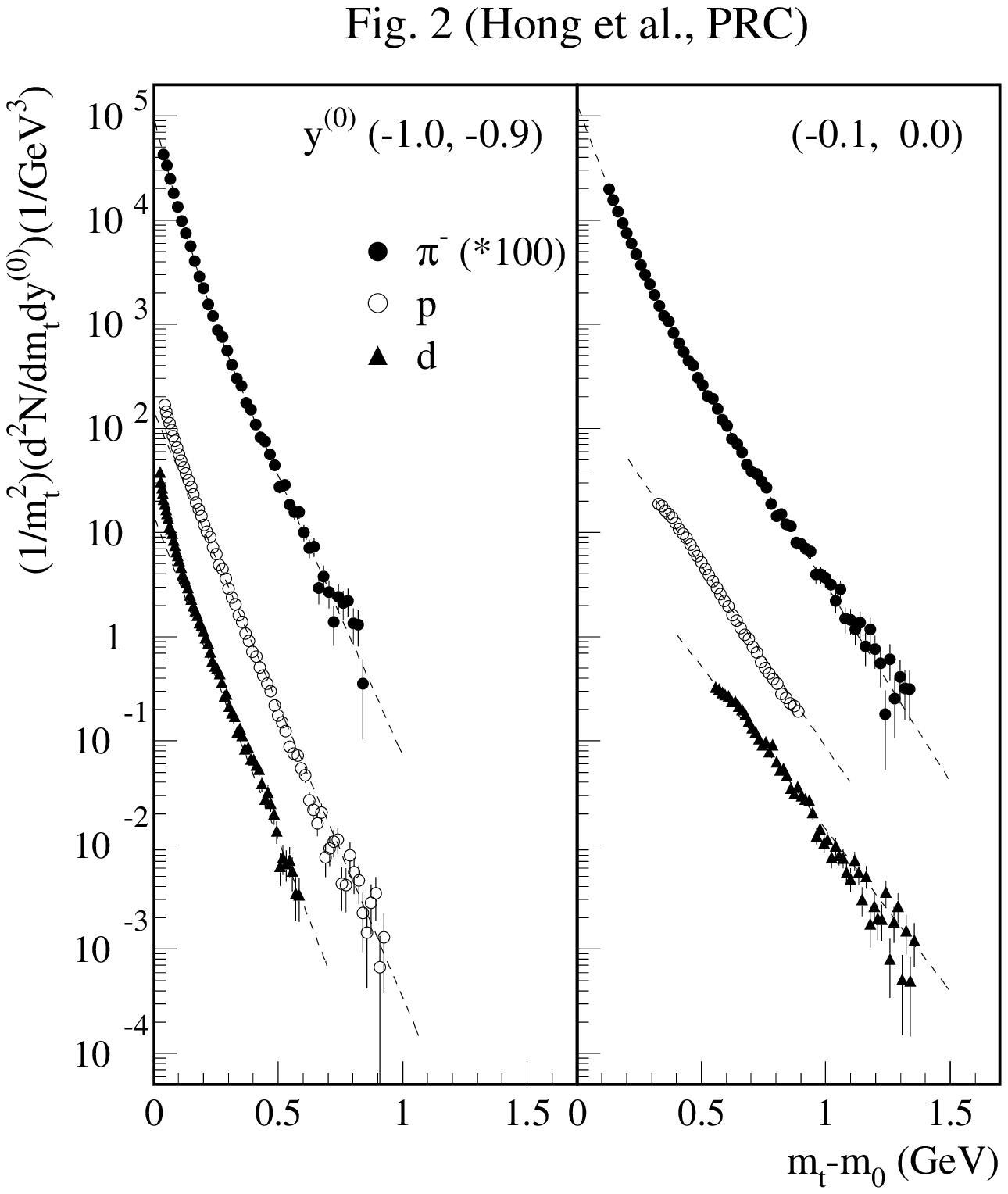}
\end{figure}

\pagebreak
\begin{figure}[hhh]
.
\includegraphics{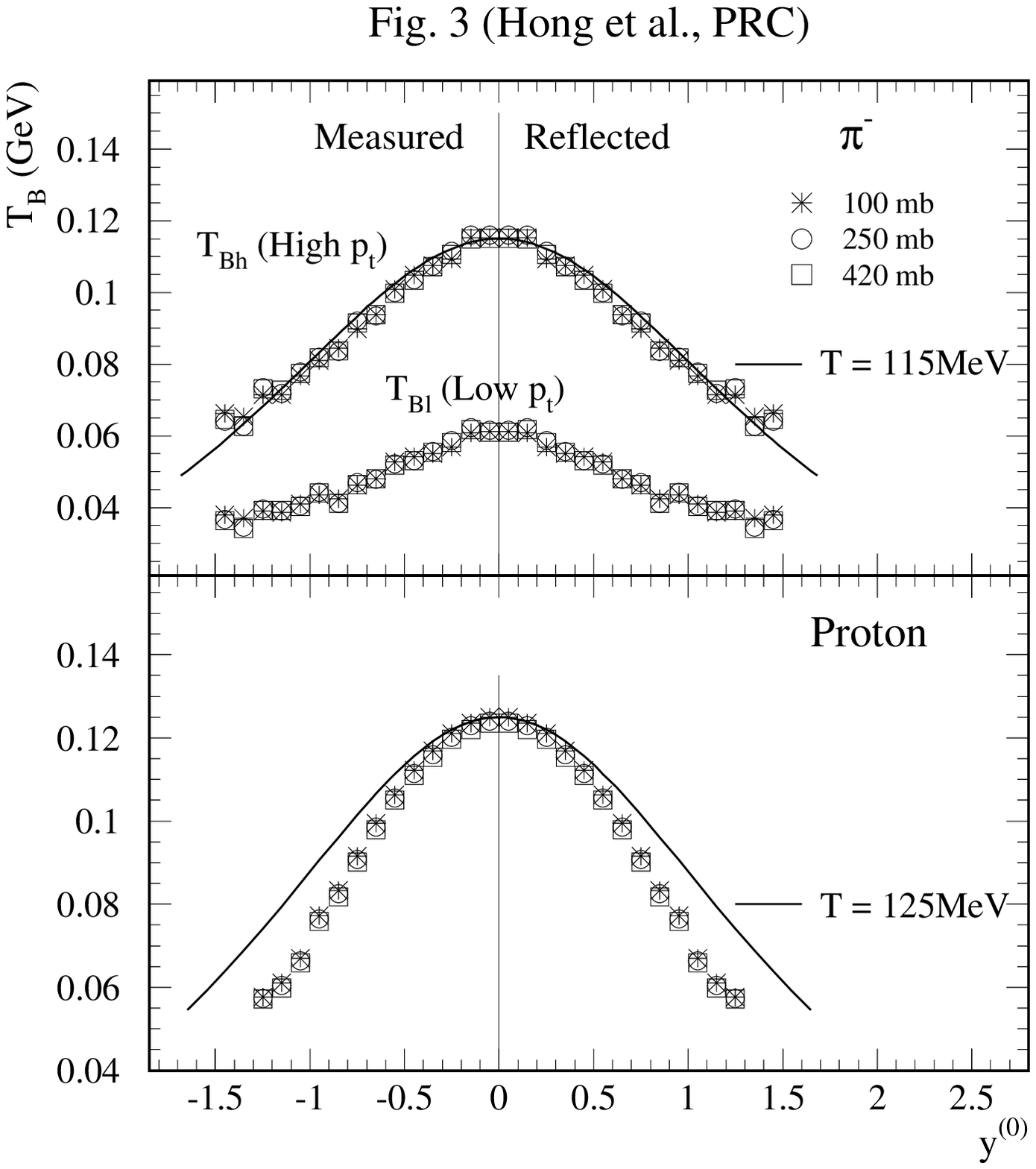}
\end{figure}

\pagebreak
\begin{figure}[hhh]
.
\includegraphics{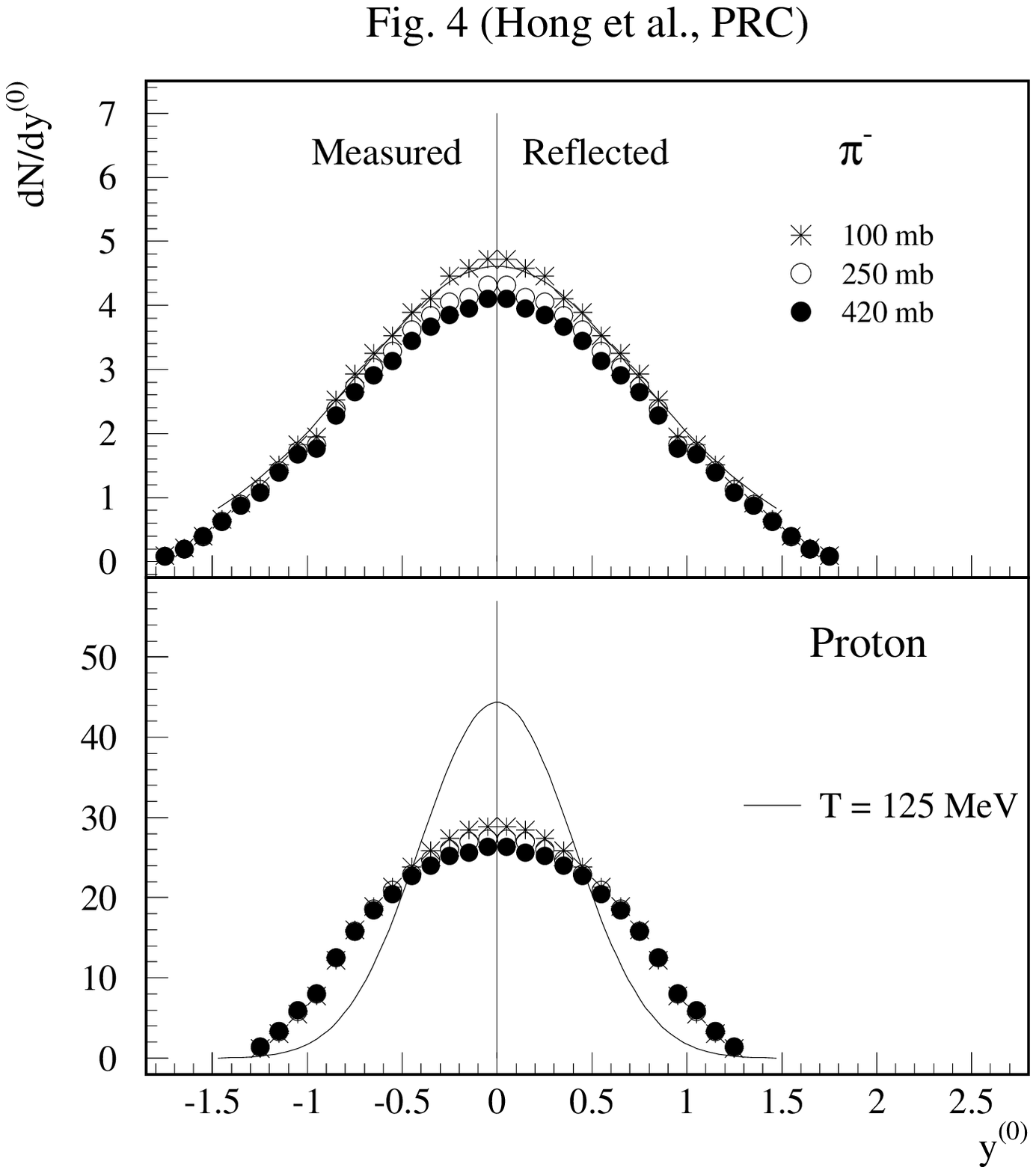}
\end{figure}

\pagebreak
\begin{figure}[hhh]
.
\includegraphics{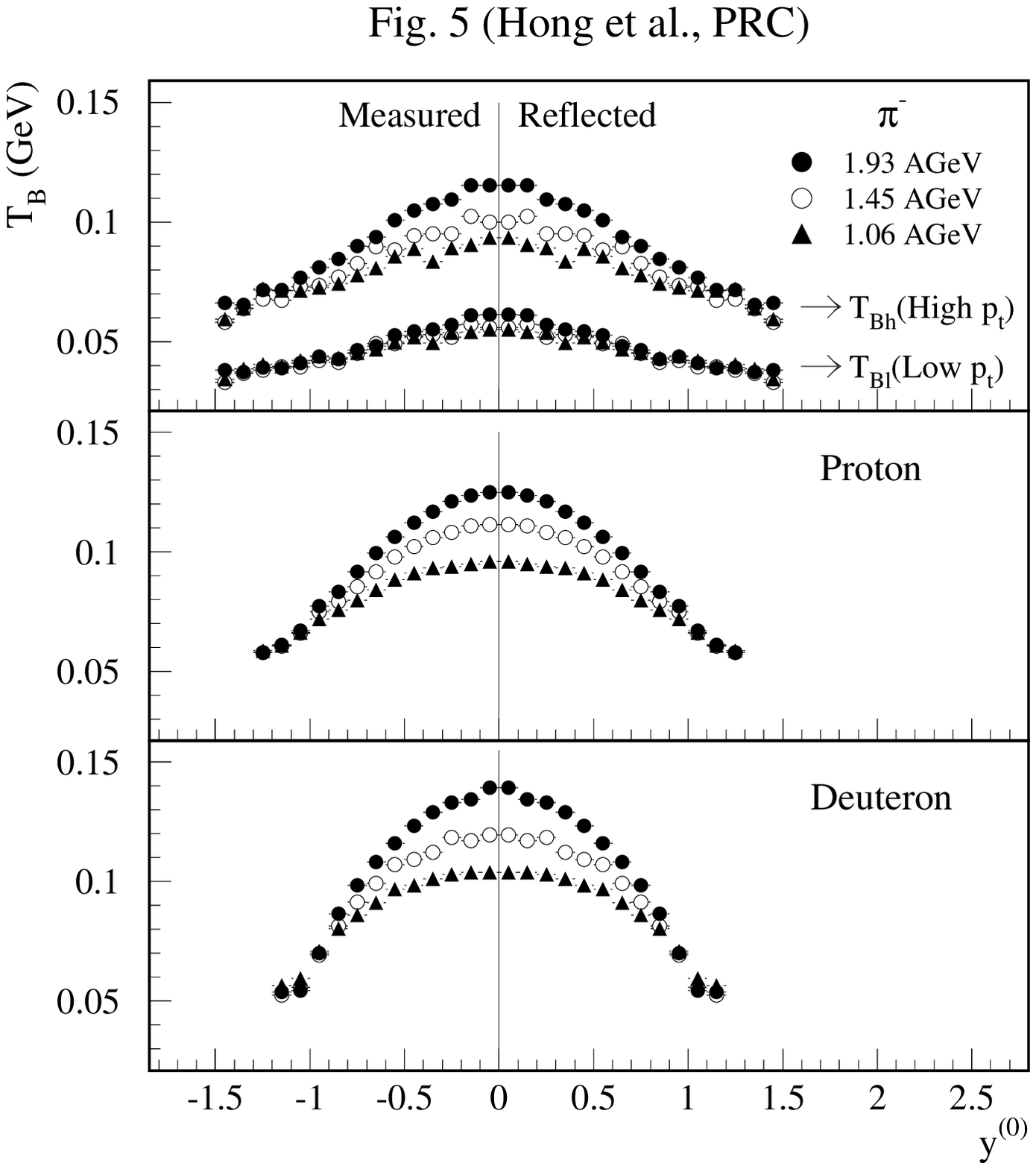}
\end{figure}

\pagebreak
\begin{figure}[hhh]
.
\includegraphics{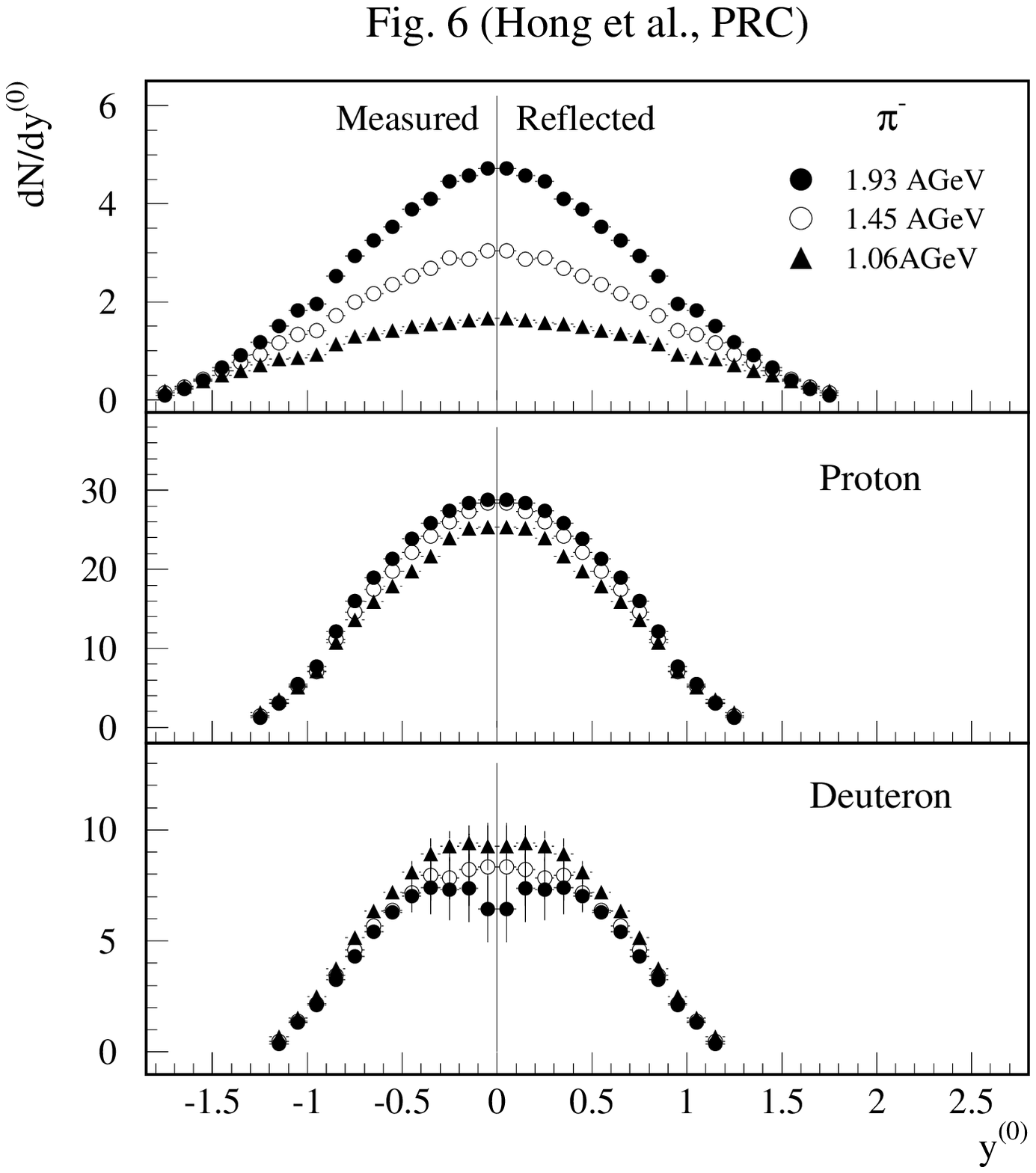}
\end{figure}

\pagebreak
\begin{figure}[hhh]
.
\includegraphics{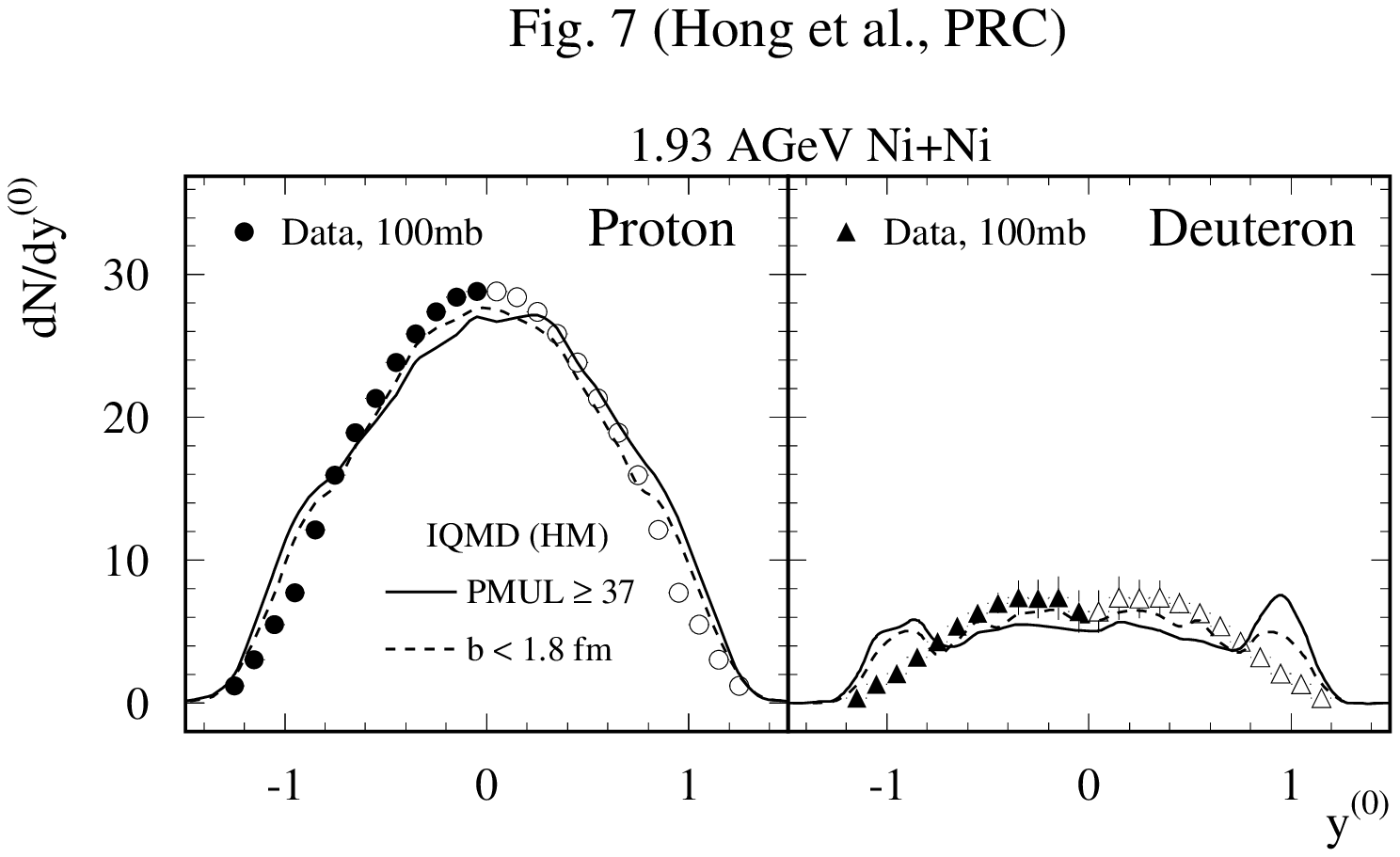}
\end{figure}

\pagebreak
\begin{figure}[hhh]
.
\includegraphics{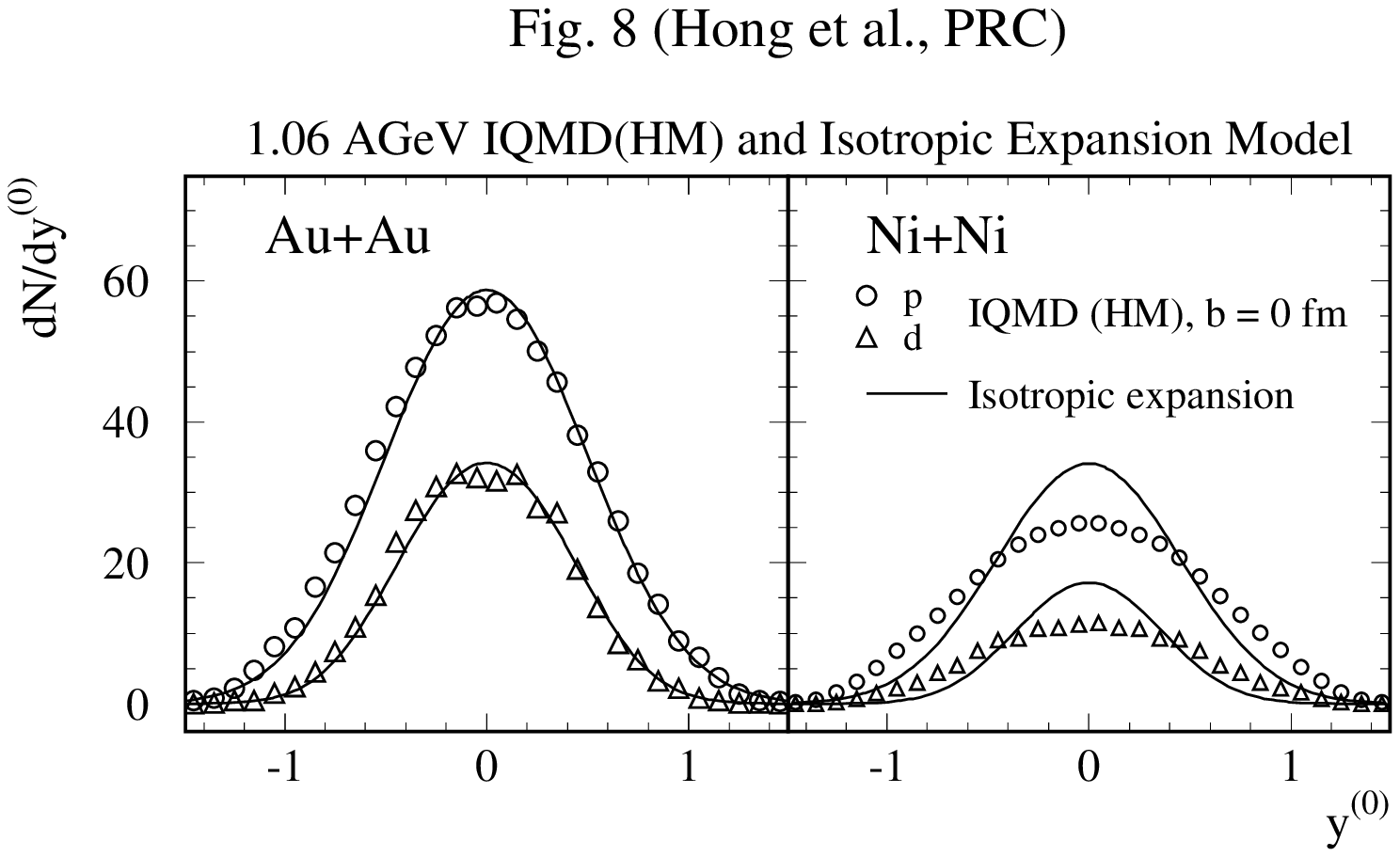}
\end{figure}

\pagebreak
\begin{figure}[hhh]
.
\includegraphics{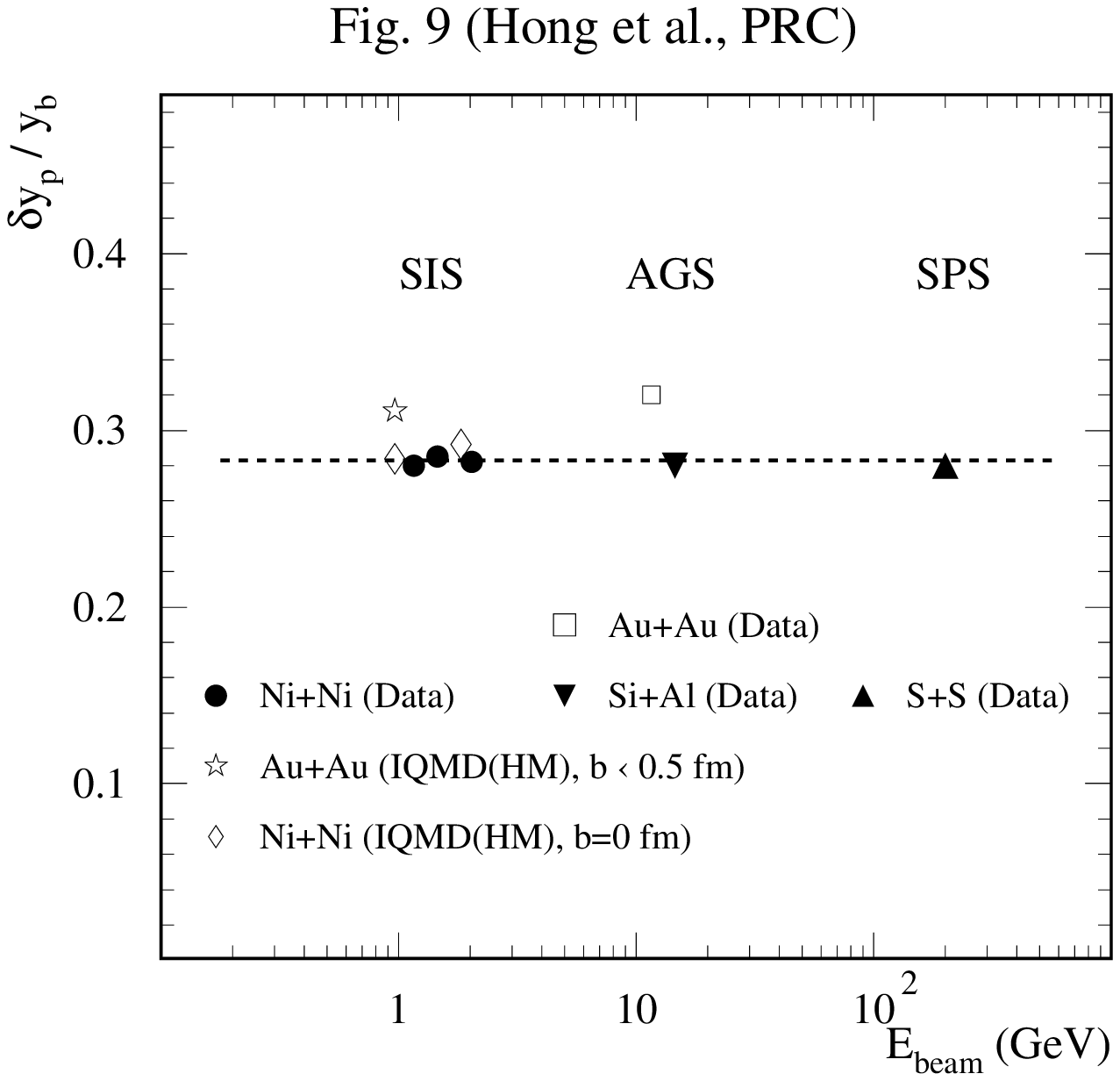}
\end{figure}

\pagebreak
\begin{figure}[hhh]
.
\includegraphics{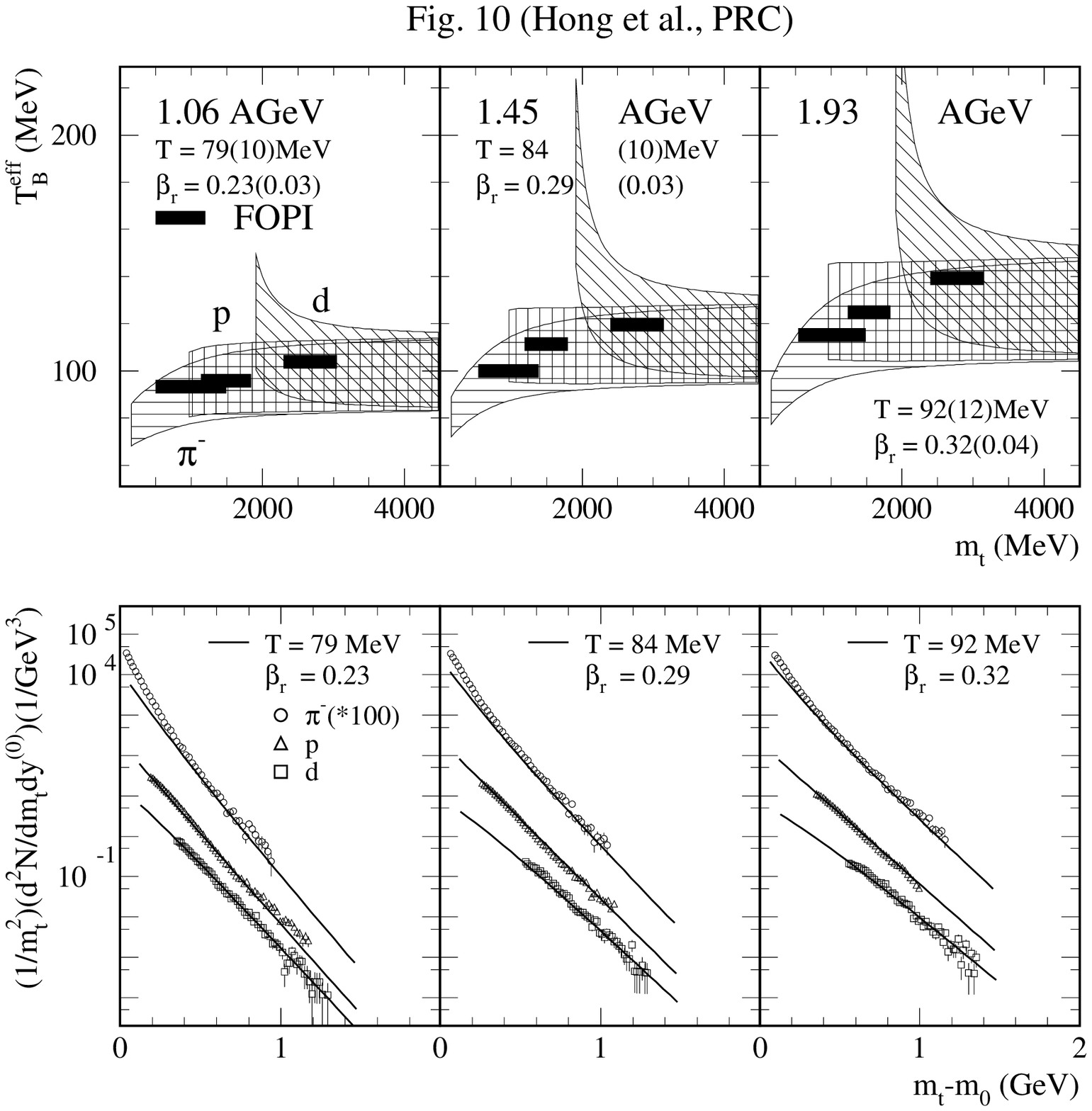}
\end{figure}

\pagebreak
\begin{figure}[hhh]
.
\includegraphics{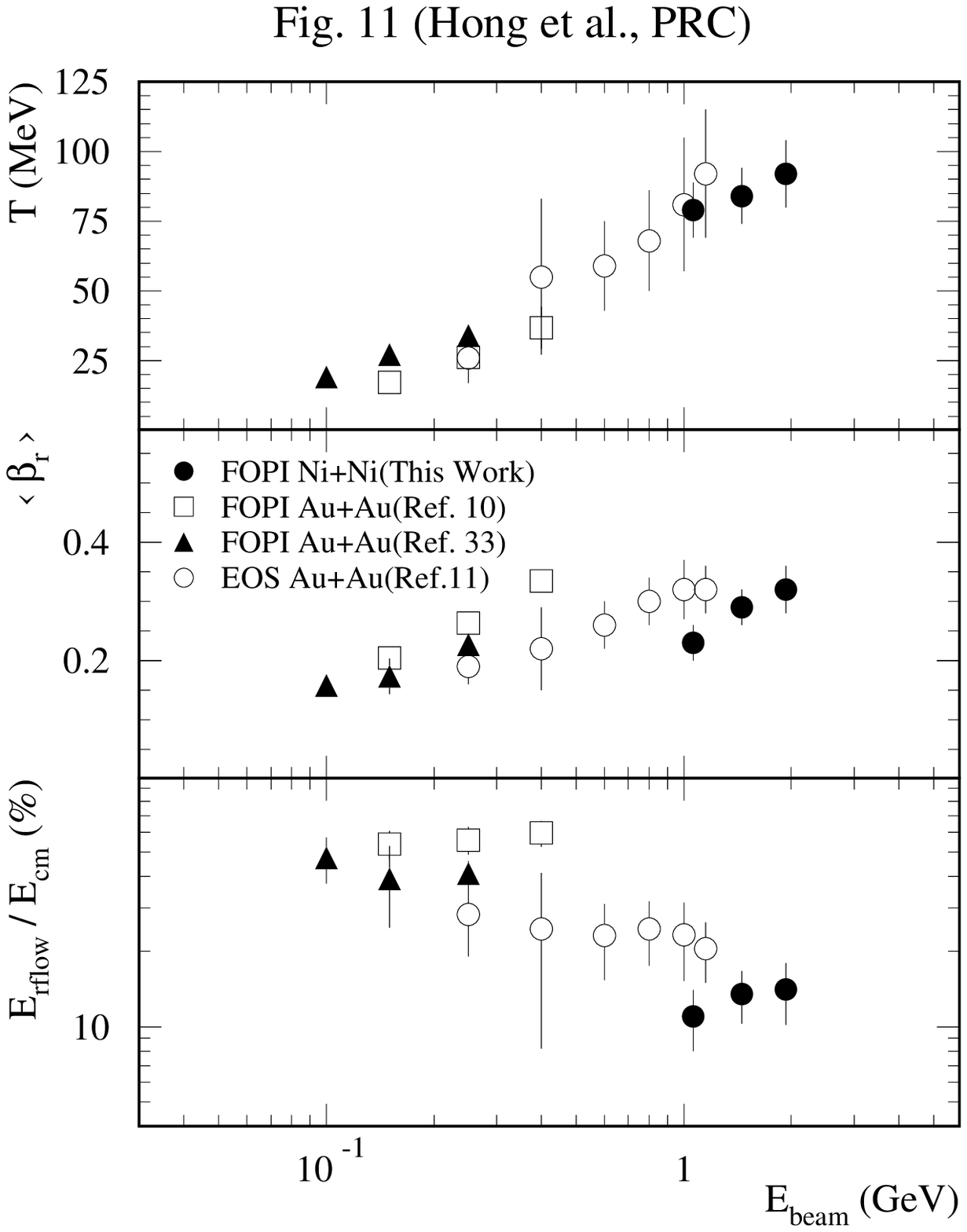}
\end{figure}

\end{document}